\documentclass[aps,preprint]{revtex4}%
\usepackage{amsfonts}
\usepackage{amsmath}
\usepackage{amssymb}
\usepackage{graphicx}%
\begin{document}
\preprint{ }
\title{Temperature-dependent errors in nuclear lattice simulations}
\author{Dean Lee and Richard Thomson}
\affiliation{Department of Physics, North Carolina State University, Raleigh, NC 27695-8202}

\begin{abstract}
\noindent We study the temperature dependence of discretization errors in
nuclear lattice simulations. \ We find that for systems with strong attractive
interactions the predominant error arises from the breaking of Galilean
invariance. \ We propose a local \textquotedblleft
well-tempered\textquotedblright\ lattice action which eliminates much of this
error. \ The well-tempered action can be readily implemented in lattice
simulations for nuclear systems as well as cold atomic Fermi systems.

\end{abstract}
\date{\today}
\maketitle

\section{Introduction}

Nuclear lattice simulations address the nuclear many-body problem by combining
numerical lattice methods with effective field theory. \ There have been
several recent studies on the subject of nuclear lattice simulations
\cite{Muller:1999cp,Abe:2003fz,Lee:2004si,Lee:2004qd,Hamilton:2004gp,Seki:2005ns,Lee:2005is,Lee:2005it,Borasoy:2005yc,Borasoy:2006qn}%
. \ The starting point is the usual starting point of effective field theory.
\ All local interactions consistent with the symmetries of low-energy nuclear
physics are organized by counting factors of $Q/\Lambda_{\text{high}}$,
where$\ Q$ is the typical nucleon momentum scale and $\Lambda_{\text{high}}$
is the high-momentum scale where the effective theory eventually breaks down.
\ For chiral effective field theory we have $\Lambda_{\text{high}}\sim4\pi
f_{\pi},$ and for effective field theory without pions $\Lambda_{\text{high}%
}\sim m_{\pi}$. \ Interactions in the effective theory are truncated at some
order in $Q/\Lambda_{\text{high}}$, and the remaining interactions are put on
a space-time lattice. \ Coefficients for the interactions are determined by
matching to scattering phase shifts and few-nucleon spectra. \ Once the
interaction coefficients of the lattice effective theory are determined, the
many-body system can be simulated nonperturbatively using Monte Carlo. \ The
method can be applied to nuclei at zero temperature as well as the
thermodynamics of nuclear and neutron matter at nonzero temperature. \ Similar
lattice effective field theory techniques have been used to study cold atomic
Fermi systems in the limit of short range interactions and long scattering
length
\cite{Chen:2003vy,Wingate:2005xy,Bulgac:2005a,Lee:2005fk,Burovski:2006a,Burovski:2006b,Lee:2006hr}%
.

In addition to $Q$ and $\Lambda_{\text{high}}$ there is also a momentum cutoff
scale $\Lambda$. \ On the lattice with lattice spacing $a$, the cutoff
momentum scale is $\Lambda=\pi a^{-1}$. \ Ideally one should increase
$\Lambda$ systematically to extrapolate to the continuum limit $\Lambda
\rightarrow\infty$. \ For any finite set of diagrams with the required local
counterterms this is not a problem. \ However when diagrams are iterated to
all orders complications arise when the interactions involve singular
potentials. \ For example in pionless effective field theory it is known that
a three-body counterterm is required at leading order
\cite{Efimov:1971a,Efimov:1993a,Bedaque:1998kg,Bedaque:1998km,Bedaque:1999ve,Platter:2004pra,Platter:2004zs,Braaten:2004a,Borasoy:2005yc,Platter:2006ad}%
.\ \ With the three-body counterterm in place the continuum limit is well
defined for few-body calculations. \ Unfortunately at very high cutoff
momentum this approach involves removing spurious deeply-bound states by hand,
and there is no way to do this in many-body simulations. \ The problem is no
better in effective field theory with pions. \ In this case the pion tensor
force generates instabilities in higher partial wave channels at large
$\Lambda$ \cite{Nogga:2005hy,Birse:2005um,Epelbaum:2005pn,Epelbaum:2006pt}.
\ In short the presence of continuum limit instabilities and computational
constraints means that for many-body simulations one is restricted to a finite
range of values for $\Lambda$. \ Therefore it is important to understand and
control errors that occur at finite cutoff momentum.

In this study we investigate finite cutoff errors on the lattice at nonzero
temperature. \ While our results and conclusions apply to general few- and
many-body nuclear systems, we center our discussion on simulations of dilute
neutron matter. \ In particular we consider an idealized limit of neutron
matter with zero range two-body interactions. \ We take this zero-range
two-body contact interaction as the only interaction. \ Therefore
$\Lambda_{\text{high}}\rightarrow\infty,$ and it is straightforward to show
that this idealized theory has no continuum limit instabilities. \ For these
reasons it is a useful testing ground to study $Q/\Lambda$ cutoff errors
without additional complications. \ Zero-range neutron matter is a good
approximation to actual dilute neutron matter when the spacing between
neutrons is sufficiently large. \ This occurs at about $1\%$ of normal nuclear
matter density or less.

Our interest in finite cutoff errors at nonzero temperature is motivated by a
recent analysis of zero-range neutron matter on the lattice which found
sizable lattice errors at nonzero temperature \cite{Lee:2005is}. \ This stands
in contrast with zero temperature simulations which found little dependence on
lattice spacing \cite{Lee:2005fk,Lee:2006hr}. \ The lattice spacing dependence
at nonzero temperature was first noticed in the results of many-body lattice
simulations and then analyzed by calculating coefficients of the virial
expansion. \ The second-order virial coefficient $b_{2}(T)$, where $T$ is
temperature, was found to be too large when computed on the lattice. \ While
the source of the error was unknown, it was suggested that tuning the two-body
interaction to give the correct value for $b_{2}(T)$ might improve the results
of the many-body simulation. \ This suggestion was carried out in
\cite{Lee:2005it}, and the many-body lattice results with the retuned
interaction showed little residual dependence on lattice spacing. \ Similar
cutoff errors were found in \cite{Burovski:2006a,Burovski:2006b}. \ However
the analysis did not distinguish between cutoff errors due to nonzero
temperature and cutoff errors due to nonzero density.

In this paper we answer some of the questions raised by the findings in
\cite{Lee:2005is} and \cite{Lee:2005it}. \ In particular we discuss the source
of the large temperature-dependent lattice errors, why the measured energies
tended to be too low, and why in \cite{Lee:2005it} it was possible to cancel
much of the error by retuning the two-body interaction. \ We also propose a
simple modified lattice action which eliminates most of the large
temperature-dependent lattice errors from the beginning. \ The results of our
analysis should be useful for reducing systematic errors in future nuclear
lattice simulations as well as other strongly-attractive fermionic systems.

\section{Virial expansion}

The virial expansion for the equation of state has been used to study neutron
and nuclear matter as well as fermionic atoms near the classical regime
\cite{Ho:2004a,Horowitz:2005nd,Horowitz:2005zv,Rupak:2006pu} \ The virial
expansion can be regarded as a power series in fugacity, $z=e^{\beta\mu}$,
where $\beta$ is the inverse temperature and $\mu$ is the chemical potential.
\ For example the logarithm of the grand canonical partition function per unit
volume for neutron matter can be written as%
\begin{equation}
\frac{1}{V}\ln Z_{G}=\frac{2}{\lambda_{T}^{3}}\left[  z+b_{2}(T)z^{2}%
+b_{3}(T)z^{3}\cdots\right]  ,
\end{equation}
where
\begin{equation}
\lambda_{T}=\sqrt{\frac{2\pi\beta}{m}}%
\end{equation}
is the thermal wavelength and $m$ is the neutron mass. \ We can use the virial
expansion to compute thermodynamic observables when the thermal wavelength is
smaller than the interparticle spacing. \ The neutron density, $\rho$, can be
computed by taking a derivative of $\ln Z_{G}$ with respect to the chemical
potential,
\begin{equation}
\rho=\frac{1}{\beta V}\frac{\partial}{\partial\mu}\ln Z_{G}\text{.}%
\end{equation}
To second order in the virial expansion we find%
\begin{equation}
\rho=\frac{2}{\lambda_{T}^{3}}\left[  z+2b_{2}(T)z^{2}+\cdots\right]  .
\label{virialrho}%
\end{equation}
Taking into account Fermi statistics, we get%
\begin{equation}
b_{2}^{\text{free}}(T)=-2^{-5/2}\approx-0.177
\end{equation}
for a free gas of neutrons.

With the interactions turned on, the second virial coefficient can be computed
by extracting the term in the partition function proportional to $z^{2}$,%
\begin{equation}
b_{2}(T)-b_{2}^{\text{free}}(T)=\frac{\lambda_{T}^{3}}{2V}\left\{
Tr_{2}[\text{exp}(-\beta H)]-Tr_{2}[\text{exp}(-\beta H_{\text{free}%
})]\right\}  . \label{traceb2}%
\end{equation}
$Tr_{2}$ denotes the trace over all two neutron states, $H$ is the full
Hamiltonian, and $H_{\text{free}}$ is the free Hamiltonian. \ By integrating
over the center of mass momentum and enforcing spherical boundary conditions
on the relative displacement between the two particles, the density of
scattering states can be related to the total elastic phase shift $\delta(E)$
\cite{Beth:1937},%

\begin{equation}
b_{2}(T)-b_{2}^{\text{free}}(T)=\frac{\beta}{2^{1/2}\pi}\int_{0}^{\infty
}dE\;e^{-\beta E/2}\delta(E)+\text{bound state contribution.}%
\end{equation}
If there are two-body bound states in the spectrum with binding energies
$E_{B,i}$, there is an additional contribution%
\begin{equation}
\frac{3}{2^{1/2}}%
{\displaystyle\sum\limits_{i}}
\left(  e^{\beta\left\vert E_{B,i}\right\vert }-1\right)  .
\end{equation}
In the unitary limit, where the effective range is zero and scattering length
is infinite, we get%
\begin{equation}
b_{2}(T)=3\times2^{-\frac{5}{2}}\approx0.530.
\end{equation}
For zero effective range but arbitrary scattering length $a_{\text{scatt}}$
the second virial coefficient is%
\begin{equation}
b_{2}(T)=\left\{
\genfrac{}{}{0pt}{}{\frac{e^{x^{2}}}{\sqrt{2}}\left[  1-\operatorname{erf}%
(\left\vert x\right\vert )\right]  -\frac{1}{4\sqrt{2}}\text{ \ for
}x<0,}{\sqrt{2}e^{\frac{\left\vert E_{B}\right\vert }{k_{B}T}}-\frac{e^{x^{2}%
}}{\sqrt{2}}\left[  1-\operatorname{erf}(x)\right]  -\frac{1}{4\sqrt{2}}\text{
\ for }x>0,}%
\right.
\end{equation}
where $\operatorname{erf}$ is the error function, $E_{B}$ is the two-particle
binding energy for positive scattering length, and%
\begin{equation}
x=\frac{\lambda_{T}}{\sqrt{2\pi}a_{\text{scatt}}}.
\end{equation}
As the effective range goes to zero we have the relation%
\begin{equation}
\left\vert E_{B}\right\vert =\frac{1}{ma_{\text{scatt}}^{2}},
\end{equation}
and therefore we can write%
\begin{equation}
b_{2}(T)=\left\{
\genfrac{}{}{0pt}{}{\frac{e^{x^{2}}}{\sqrt{2}}\left[  1-\operatorname{erf}%
(\left\vert x\right\vert )\right]  -\frac{1}{4\sqrt{2}}\text{ \ for
}x<0,}{\frac{e^{x^{2}}}{\sqrt{2}}\left[  1+\operatorname{erf}(x)\right]
-\frac{1}{4\sqrt{2}}\text{ \ for }x>0.}%
\right.  \label{b2}%
\end{equation}

\section{One-dimensional model}

We begin our analysis of finite cutoff errors with a one-dimensional model.
\ The model consists of nonrelativistic spin-1/2 fermions with an attractive
zero-range interaction and is the continuum limit of the attractive
one-dimensional Hubbard model. \ Both the attractive and repulsive versions of
the one-dimensional Hubbard model have been studied in the literature
\cite{Yang:1967bm,Lieb:1968PRL,McGuire:1966,Ahn:1994vh,Matveenko:1997,Ortiz:1999,Salwen:2003}%
. \ We consider the attractive case as a toy model of short-range attractive
forces in nuclear systems. \ As we will see, the problem of large
discretization errors appears even in this one-dimensional model which has no
ultraviolet divergences.

\begin{figure}
[ptb]
\begin{center}
\includegraphics[
height=1.2104in,
width=5.0286in
]%
{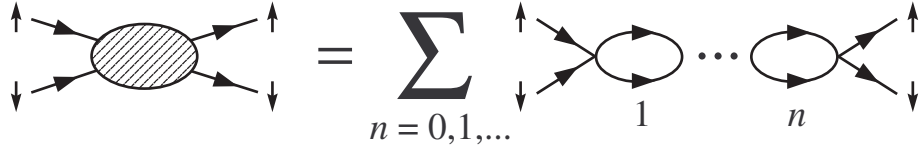}%
\caption{The connected amputated two-particle Green's function.}%
\label{twotwo}%
\end{center}
\end{figure}

In the continuum limit the Hamiltonian has the form%
\begin{equation}
H=-\frac{1}{2m}\sum_{i=\uparrow,\downarrow}\int dx\;a_{i}^{\dagger}%
(x)\frac{\partial^{2}}{\partial x^{2}}a_{i}(x)+C\int dx\;a_{\downarrow
}^{\dagger}(x)a_{\uparrow}^{\dagger}(x)a_{\uparrow}(x)a_{\downarrow}(x),
\label{continuum1D}%
\end{equation}
where $m$ is the mass, $C<0$, and $a_{i}$ and $a_{i}^{\dagger}$ are
annihilation and creation operators for spin $i$. \ The connected amputated
two-particle Green's function equals the sum of bubble diagrams shown in Fig.
\ref{twotwo}.%
Any connected scattering process consists of two-particle Green's functions
linked together with free particle propagators.

Let $G_{2}(p_{0},p_{x})$ be the amplitude for the connected amputated
two-particle Green's function, where$\ p_{0}$ is the total energy and $p_{x}$
is the total spatial momentum of the two particles. \ We sum the bubble
diagrams in Fig. \ref{twotwo} and get%
\begin{equation}
G_{2}(p_{0},p_{x})=\frac{-iC}{1-iC\cdot\Pi\left(  p_{0},p_{x}\right)  }%
=\frac{1}{-\frac{1}{iC}+\Pi\left(  p_{0},p_{x}\right)  },
\end{equation}
where
\begin{equation}
\Pi\left(  p_{0},p_{x}\right)  =\int\frac{dq_{0}dq_{x}}{\left(  2\pi\right)
^{2}}\frac{i}{\frac{p_{0}}{2}+q_{0}-\frac{\left(  \frac{p_{x}}{2}%
+q_{x}\right)  ^{2}}{2m}+i\varepsilon}\times\frac{i}{\frac{p_{0}}{2}%
-q_{0}-\frac{\left(  \frac{p_{x}}{2}-q_{x}\right)  ^{2}}{2m}+i\varepsilon}.
\label{onebubble}%
\end{equation}
In the continuum limit we find%
\begin{align}
\Pi\left(  p_{0},p_{x}\right)   &  =-\frac{m}{2\sqrt{mp_{0}-\frac{p_{x}^{2}%
}{4}}},\\
G_{2}(p_{0},p_{x})  &  =\frac{1}{-\frac{1}{iC}-\frac{m}{2\sqrt{mp_{0}%
-\frac{p_{x}^{2}}{4}}}}.
\end{align}
Since $C<0$ there is a bound state pole in the two-particle Green's function
at energy%
\begin{equation}
p_{0}=-\frac{mC^{2}}{4}+\frac{p_{x}^{2}}{4m}.
\end{equation}
We can obtain the same result by solving the Schr\"{o}dinger equation for the
two-particle system in the center of mass frame. \ If $x$ is the relative
separation between particles then, in the center of mass frame, the reduced
Hamiltonian is%
\begin{equation}
H_{CM}=-\frac{1}{m}\frac{\partial^{2}}{\partial x^{2}}+C\delta(x),
\end{equation}
and the ground state wavefunction is%
\begin{equation}
\psi_{0}(x)\propto\exp\left(  \frac{1}{2}mC\left\vert x\right\vert \right)  .
\label{wavefunction}%
\end{equation}
The ground state energy with center of mass kinetic energy included is%
\begin{equation}
E_{0}(p_{x})=-\frac{mC^{2}}{4}+\frac{p_{x}^{2}}{4m},
\end{equation}
where $p_{x}$ is the total momentum.

The one-dimensional system is finite in the continuum limit and therefore no
regularization nor renormalization is needed. \ Nevertheless we impose an
ultraviolet cutoff on the momentum in order to study the resulting cutoff
errors. \ \ With a momentum cutoff at $\Lambda$ we find%
\begin{equation}
\Pi\left(  p_{0},p_{x},\Lambda\right)  =-\frac{m}{2\sqrt{mp_{0}-\frac
{p_{x}^{2}}{4}}}\times\left[  1+O\left(  \frac{Q^{2}}{\Lambda^{2}}\right)
\right]  , \label{onebubble_lambda}%
\end{equation}
where $Q^{2}$ is some homogeneous combination of the parameters $mp_{0}$ and
$p_{x}^{2}$. \ The combination will depend on the details of the chosen
regularization scheme. \ The regularized two-particle Green's function has the
form%
\begin{equation}
G_{2}(p_{0},p_{x},\Lambda)=\frac{1}{-\frac{1}{iC(\Lambda)}-\frac{m}%
{2\sqrt{mp_{0}-\frac{p_{x}^{2}}{4}}}\times\left[  1+O\left(  \frac{Q^{2}%
}{\Lambda^{2}}\right)  \right]  }. \label{1D_regularized_Greens}%
\end{equation}
We define the scale-dependent coupling $C(\Lambda)$ so that the pole in the
rest frame remains exactly at%
\begin{equation}
p_{0}=-\frac{mC^{2}}{4}.
\end{equation}

\subsection{One- and two-particle lattice dispersion relation in one
dimension}

We investigate the cutoff error in more detail using a Hamiltonian lattice
formalism. \ On the lattice the cutoff momentum scale $\Lambda$ corresponds
with $\pi a^{-1}$, where $a$ is the lattice spacing. \ Throughout our
discussion of the lattice formalism we use dimensionless parameters and
operators, which correspond with physical values multiplied by the appropriate
power of $a$. \ However final results are reported into physical units. \ We
start with the simplest possible lattice Hamiltonian giving (\ref{continuum1D}%
) in the continuum limit. \ We let%
\begin{equation}
H^{(0)}=K^{(0)}+V^{(0)},
\end{equation}%
\begin{equation}
K^{(0)}=\frac{1}{m}\sum_{n_{x},i}a_{i}^{\dagger}(n_{x})a_{i}(n_{x})-\frac
{1}{2m}\sum_{n_{x},i}\left[  a_{i}^{\dagger}(n_{x})a_{i}(n_{x}+1)+a_{i}%
^{\dagger}(n_{x})a_{i}(n_{x}-1)\right]  , \label{K0}%
\end{equation}%
\begin{equation}
V^{(0)}=C\sum_{n_{x}}a_{\downarrow}^{\dagger}(n_{x})a_{\uparrow}^{\dagger
}(n_{x})a_{\uparrow}(n_{x})a_{\downarrow}(n_{x}). \label{V0}%
\end{equation}
We refer to $H^{(0)}$ as the standard lattice Hamiltonian. \ The zero
superscript signifies that it is the simplest possible lattice formulation.
\ Later in our discussion we consider more complicated lattice actions. \ We
choose the mass to be $m=939$ MeV and fix the lattice spacing at $a=(50$
MeV$)^{-1}$. \ This corresponds with a cutoff momentum of $\Lambda=\pi
a^{-1}\simeq157$ MeV.

\begin{figure}
[ptb]
\begin{center}
\includegraphics[
height=2.5244in,
width=3.5276in
]%
{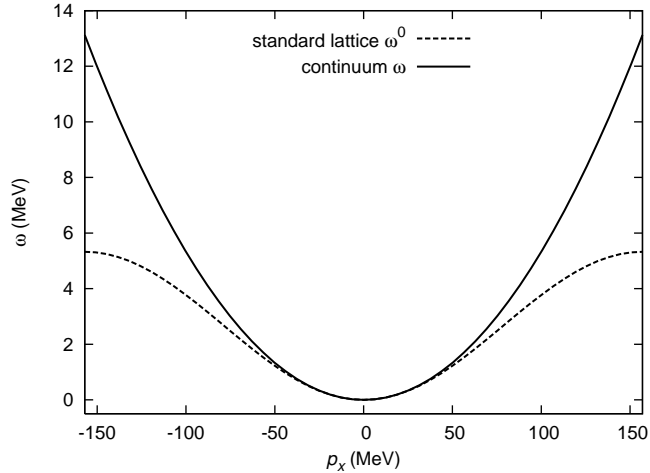}%
\caption{Single-particle dispersion relation for the standard lattice action
with $m=939$ MeV and $\Lambda=\pi a^{-1}\simeq157$ MeV. \ We also show the
continuum limit.}%
\label{dispersion1}%
\end{center}
\end{figure}

The single-particle dispersion relation for the standard lattice Hamiltonian
is given by%
\begin{equation}
\omega^{(0)}(p_{x})=\frac{1}{m}\times\left(  1-\cos p_{x}\right)  =\frac
{p_{x}^{2}}{2m}+O\left(  p_{x}^{4}\right)  . \label{disprel0}%
\end{equation}
In Fig. \ref{dispersion1} we have plotted $\omega^{(0)}(p_{x})$ versus the
continuum result $\omega(p_{x})=p_{x}^{2}/(2m)$ for momenta in the first
Brillouin zone $\left\vert p_{x}\right\vert \leq\Lambda$.%
The relative error between $\omega^{(0)}$ and $\omega$ is $10\%$ or less for
$\left\vert p_{x}\right\vert <\Lambda/3.$

We now consider two-particle states with one spin-up particle and one
spin-down particle. \ We start with a small value for the coupling,
$C=-0.0400$. \ Let $p_{x}$ be the total momentum of the two-particle system.
\ In the continuum limit the ground state energy at zero total momentum is%
\begin{equation}
E_{0}(p_{x}=0)=-\frac{mC^{2}}{4}=-0.3756\text{\ MeV.}%
\end{equation}
It is convenient to place the two-particle system in a periodic box. \ We
choose the box length to be $L=1$ MeV$^{-1}$. \ Since the ground state
wavefunction depends on the relative separation $x$ as%
\begin{equation}
\psi_{0}(x)\propto\exp\left(  \frac{1}{2}mC\left\vert x\right\vert \right)
\approx\exp\left[  -(19\text{\ MeV)}\cdot\left\vert x\right\vert \right]  ,
\end{equation}
the effect of the boundary at $L=1$ MeV$^{-1}$ on the ground state energy is
negligible. \ The box length does however determine the level spacing between
unbound scattering states.

\begin{figure}
[ptb]
\begin{center}
\includegraphics[
height=2.5244in,
width=3.5276in
]%
{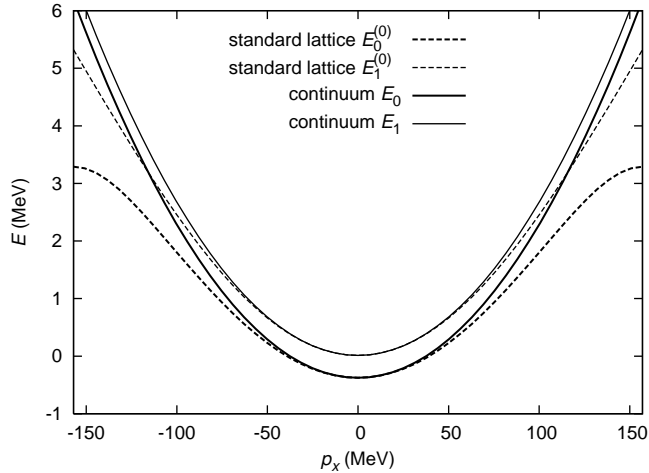}%
\caption{The lowest two-particle energies for the standard lattice action,
$E_{0}^{(0)}$ and $E_{1}^{(0)}$, and the corresponding continuum limit values,
$E_{0}$ and $E_{1}$. \ The continuum coupling is $C=-0.0400$ while the lattice
coupling is $C(\Lambda)=-0.0407.$}%
\label{standard_weak_1d}%
\end{center}
\end{figure}

For the lattice calculation the interaction strength is tuned so that the
ground state energy in the rest frame matches the continuum result of
$-0.3756$\ MeV. \ This gives an adjusted coefficient of $C(\Lambda
)\simeq-0.0407.$ \ In Fig. \ref{standard_weak_1d} we show the two lowest
energy levels of the two-particle system as functions of $p_{x}$. \ The
two-particle ground state and lowest scattering state energies for the lattice
are labelled $E_{0}^{(0)}$ and $E_{1}^{(0)}$ respectively, while the
corresponding continuum limit values are labelled $E_{0}$ and $E_{1}$.%
The mismatch between lattice and continuum results for the two-particle
energies is roughly the same size as the mismatch between single-particle
lattice and continuum kinetic energies, $\omega^{(0)}$ and $\omega$.

\begin{figure}
[ptb]
\begin{center}
\includegraphics[
height=2.5244in,
width=3.5276in
]%
{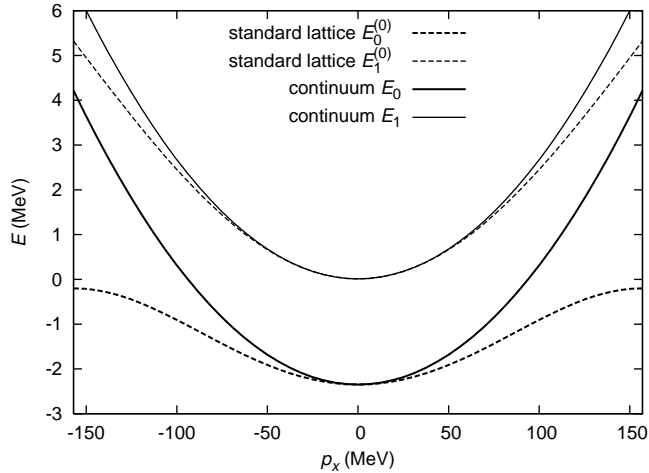}%
\caption{The lowest two-particle energies for the standard lattice action,
$E_{0}^{(0)}$ and $E_{1}^{(0)}$, and the corresponding continuum limit values,
$E_{0}$ and $E_{1}$. \ The continuum coupling is $C=-0.1000$ while the lattice
coupling is $C(\Lambda)=-0.1105.$}%
\label{standard_1d}%
\end{center}
\end{figure}

Keeping other parameters the same we now repeat the two-particle energy
calculations at stronger coupling, $C=-0.1000$. \ In this case the continuum
limit ground state energy at $p_{x}=0$ is%
\begin{equation}
E_{0}(p_{x}=0)=-\frac{mC^{2}}{4}=-2.348\text{\ MeV.}%
\end{equation}
Tuning the lattice interaction to match this ground state energy gives an
adjusted coefficient of $C(\Lambda)\simeq-0.1105$. \ Results for the
two-particle ground state and lowest scattering state are shown in Fig.
\ref{standard_1d}.%
While the agreement for the first excited states $E_{1}^{(0)}$ and $E_{1}$ has
not changed noticeably, the deviation between lattice and continuum results
for the ground state energy has increased substantially for $p_{x}\neq0$. \ We
have chosen the lattice coupling $C(\Lambda)$ so that $E_{0}^{(0)}=E_{0}$ at
$p_{x}=0,$ and so the disagreement between $E_{0}^{(0)}$ and $E_{0}$ is
proportional $p_{x}^{2}.$

\subsection{Broken Galilean invariance}

In the continuum limit Galilean invariance requires that the total energy of
the two-particle system rises quadratically with the total momentum $p_{x}$,%
\begin{equation}
E(p_{x})=E(0)+\frac{p_{x}^{2}}{4m}.
\end{equation}
However any regularization scheme with a preferred reference frame breaks
Galilean invariance to some extent. \ In the following we show how broken
Galilean invariance on the lattice can result in large cutoff errors at strong coupling.

Let the momenta of the two particles be $p_{x}/2-q_{x}$ and $p_{x}/2+q_{x}$,
where $p_{x}$ is the total momentum and $2q_{x}$ is the relative momentum
between the particles. \ We consider first the ground state. \ The average
value of the relative momentum $2q_{x}$ for the two-body ground state
wavefunction grows proportionally with $m\left\vert C\right\vert $. \ For
$C=-0.0400$ we have $m\left\vert C\right\vert =37.6$ MeV, while for
$C=-0.1000$ we have $m\left\vert C\right\vert =93.9$ MeV. \ For the latter
case $2q_{x}$ is not small compared with the cutoff momentum $157$\ MeV. \ It
is not so large as to invalidate the assumption that we have a sensible
low-energy effective field theory. \ However it is large enough that one of
the constituent particle momenta can reach the Brillouin zone boundary at
$\pm\Lambda$ even though $\left\vert p_{x}/2\right\vert $ is less than
$\Lambda$. \ At the zone boundary the lattice kinetic energy $\omega^{(0)}$ is
significantly lower than the continuum kinetic energy $\omega$. \ This error
in the dispersion relation produces a ground state energy $E_{0}^{(0)}(p_{x})$
which is lower than the continuum result $E_{0}(p_{x})$ at strong coupling.
\ This is the effect we observe in Fig. \ref{standard_1d}.

The problem with large relative momentum does not occur for low-energy
scattering states above the ground state. \ This is because the wavefunctions
for these scattering states are peaked around the asymptotic momenta of the
particles, $p_{x}/2-q_{x}$ and $p_{x}/2+q_{x}$. \ In the infinite $L$ limit we
have%
\begin{equation}
E=\frac{1}{2m}\left(  \frac{p_{x}}{2}-q_{x}\right)  ^{2}+\frac{1}{2m}\left(
\frac{p_{x}}{2}+q_{x}\right)  ^{2}=\frac{p_{x}^{2}}{4m}+\frac{q_{x}^{2}}{m},
\end{equation}%
\begin{equation}
q_{x}^{2}=mE-\frac{p_{x}^{2}}{4}.
\end{equation}
If $mE$ and $p_{x}^{2}$ are much less than $\Lambda^{2}$, it follows that
$q_{x}^{2}$ is also much less than $\Lambda^{2}$. \ Hence the single-particle
momenta $p_{x}/2-q_{x}$ and $p_{x}/2+q_{x}$ are small compared with $\Lambda,$
and cutoff errors should remain small for low-energy scattering states even at
strong coupling.

\subsection{Cutoff errors at nonzero temperature}

\begin{figure}
[ptb]
\begin{center}
\includegraphics[
height=2.5238in,
width=3.5276in
]%
{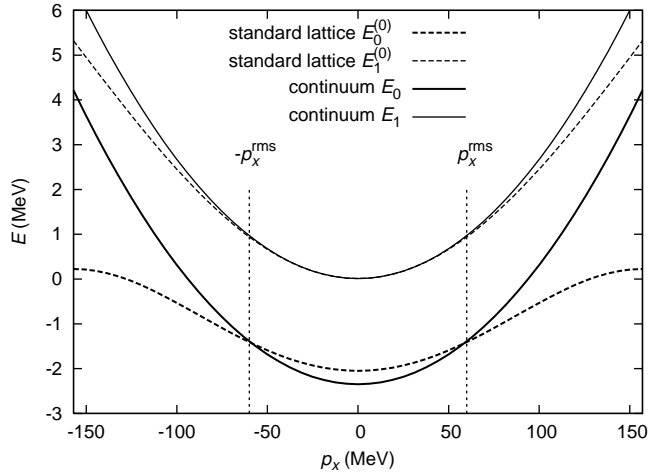}%
\caption{The two-particle energies for the lattice, $E_{0}^{(0)}$ and
$E_{1}^{(0)}$, and continuum, $E_{0}$ and $E_{1}$. The continuum coupling is
$C=-0.1000$ while the lattice coupling $C(\Lambda)$ is set by matching
$E_{0}(p_{x})$ at $\left\vert p_{x}\right\vert =p_{x}^{\text{rms}}\simeq60$
MeV.}%
\label{standard_1d_shift}%
\end{center}
\end{figure}

In the classical regime the equipartition theorem tells us that the
distribution of momenta $p_{x}$ satisfies%
\begin{equation}
\left\langle \frac{p_{x}^{2}}{4m}\right\rangle =\frac{1}{2}T,\qquad
p_{x}^{\text{rms}}=\sqrt{2mT}.
\end{equation}
Since the cutoff errors are proportional to $p_{x}^{2}$, this suggests that at
fixed lattice spacing the cutoff errors for the dilute system should increase
linearly with $T$. \ One approach to removing this error at nonzero $T$ is to
define the lattice coupling $C(\Lambda)$ by matching the continuum energy
$E_{0}(p_{x})$ at $\left\vert p_{x}\right\vert =\sqrt{2mT}$ rather than at
$p_{x}=0.$ \ In Fig. \ref{standard_1d_shift} we show the two-particle energies
when $C(\Lambda)$ is fit to $E_{0}(p_{x})$ at $\left\vert p_{x}\right\vert
=p_{x}^{\text{rms}}\simeq60$ MeV.
This redefinition has the unattractive feature that the lattice coupling
$C(\Lambda)$ is now temperature dependent. \ Furthermore it does not fix the
problem of strongly-broken Galilean invariance. \ The cutoff error has simply
been shifted to momenta $\left\vert p_{x}\right\vert \neq p_{x}^{\text{rms}}$.
\ However it does remove large cutoff errors from the lattice simulation at
temperature $T$. \ This is essentially the approach used in
\cite{Lee:2005is,Lee:2005it}, where the lattice coupling $C(\Lambda)$ was
determined by matching the continuum limit value for the second virial
coefficient $b_{2}(T)$, where $T$ is the chosen simulation temperature.

There are other techniques which actually reduce the breaking of Galilean
invariance on the lattice. \ One possibility is to remove all of the $O\left(
\frac{Q^{2}}{\Lambda^{2}}\right)  $ dependence using higher-dimensional
operators. \ This includes two-derivative interactions such as%
\begin{equation}
O_{1}(x)=\frac{d}{dx}\left[  a_{\downarrow}^{\dagger}(x)a_{\uparrow}^{\dagger
}(x)\right]  \frac{d}{dx}\left[  a_{\uparrow}(x)a_{\downarrow}(x)\right]  ,
\end{equation}%
\begin{align}
O_{2}(x)  &  =a_{\downarrow}^{\dagger}(x)a_{\uparrow}^{\dagger}(x)\left[
\frac{d^{2}a_{\uparrow}(x)}{dx^{2}}a_{\downarrow}(x)-2\frac{da_{\uparrow}%
(x)}{dx}\frac{da_{\downarrow}(x)}{dx}+a_{\uparrow}(x)\frac{d^{2}a_{\downarrow
}(x)}{dx^{2}}\right] \nonumber\\
&  +\left[  \frac{d^{2}a_{\downarrow}^{\dagger}(x)}{dx^{2}}a_{\uparrow
}^{\dagger}(x)-2\frac{da_{\downarrow}^{\dagger}(x)}{dx}\frac{da_{\uparrow
}^{\dagger}(x)}{dx}+a_{\downarrow}^{\dagger}(x)\frac{d^{2}a_{\uparrow
}^{\dagger}(x)}{dx^{2}}\right]  a_{\uparrow}(x)a_{\downarrow}(x).
\end{align}
$O_{1}$ could be tuned to cancel the broken Galilean invariance while $O_{2}$
could be tuned to reset the effective range to zero. \ However new
interactions such as these can introduce sign oscillations and other
complications in Monte Carlo simulations. \ Therefore we first try a less
expensive approach where the interaction is left alone and only the lattice
kinetic energy is modified.

\subsection{$O(a^{2})$-improved and $O(a^{2})$-well-tempered actions in one
dimension}

\begin{figure}
[ptb]
\begin{center}
\includegraphics[
height=2.5244in,
width=3.5276in
]%
{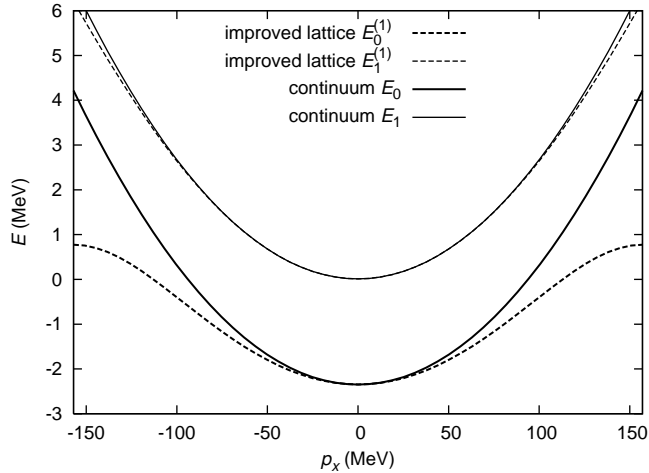}%
\caption{The lowest two-particle energies for the improved lattice action,
$E_{0}^{(1)}$ and $E_{1}^{(1)}$, and the continuum energies, $E_{0}$ and
$E_{1}$. \ The continuum coupling is $C=-0.1000$ while the lattice coupling is
$C(\Lambda)=-0.1173$.}%
\label{improved_1d}%
\end{center}
\end{figure}

Let us consider replacing the standard lattice kinetic energy action in
(\ref{K0}) with an $O(a^{2})$-improved kinetic energy with next-to-nearest
neighbor hopping,%
\begin{align}
K^{(1)}  &  =\frac{5}{4}\times\frac{1}{m}\sum_{n_{x},i}a_{i}^{\dagger}%
(n_{x})a_{i}(n_{x})-\frac{4}{3}\times\frac{1}{2m}\sum_{n_{x},i}\left[
a_{i}^{\dagger}(n_{x})a_{i}(n_{x}+1)+a_{i}^{\dagger}(n_{x})a_{i}%
(n_{x}-1)\right] \nonumber\\
&  +\frac{1}{12}\times\frac{1}{2m}\sum_{n_{x},i}\left[  a_{i}^{\dagger}%
(n_{x})a_{i}(n_{x}+2)+a_{i}^{\dagger}(n_{x})a_{i}(n_{x}-2)\right]  .
\label{K2}%
\end{align}
This gives the dispersion relation%
\begin{equation}
\omega^{(1)}(p_{x})=\frac{1}{m}\times\left[  \frac{5}{4}-\frac{4}{3}\cos
p_{x}+\frac{1}{12}\cos\left(  2p_{x}\right)  \right]  =\frac{p_{x}^{2}}%
{2m}+O\left(  p_{x}^{6}\right)  . \label{disprel1}%
\end{equation}
Matching $E_{0}(p_{x}=0)=-2.348$\ MeV for the improved lattice action gives an
adjusted coefficient of $C(\Lambda)=-0.1173$. \ Results for the two-particle
ground state and lowest scattering state with the improved action are shown in
Fig. \ref{improved_1d}.%
We see that the deviation between lattice and continuum results for the ground
state energy has been reduced for $p_{x}\neq0$.

\begin{figure}
[ptb]
\begin{center}
\includegraphics[
height=2.5244in,
width=3.5276in
]%
{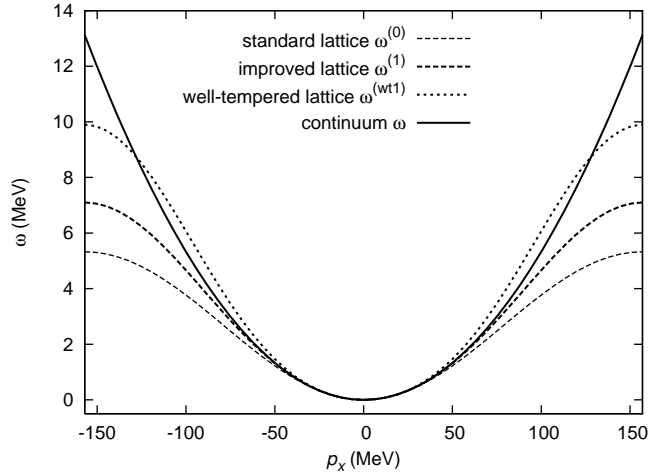}%
\caption{Comparsion of the lattice dispersion relations $\omega^{(0)}$,
$\omega^{(1)}$, $\omega^{\text{(wt1)}}$, and the continuum limit $\omega$.}%
\label{dispersion2}%
\end{center}
\end{figure}

\begin{figure}
[ptbptb]
\begin{center}
\includegraphics[
height=2.5238in,
width=3.5276in
]%
{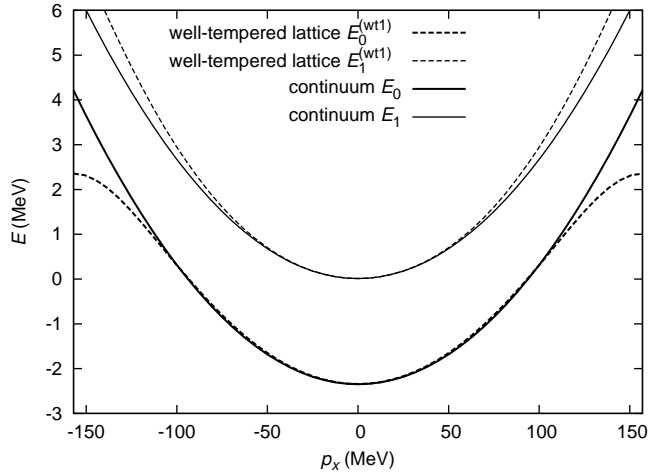}%
\caption{The lowest two-particle energies for the well-tempered lattice
action, $E_{0}^{(\text{wt}1)}$ and $E_{1}^{(\text{wt}1)}$, and continuum,
$E_{0}$ and $E_{1}$. \ The continuum coupling is $C=-0.1000$ while the lattice
coupling is $C(\Lambda)=-0.1260$.}%
\label{welltempered_1d}%
\end{center}
\end{figure}

While the errors for the improved action are better than that for the standard
action, better agreement seems possible. \ Instead of removing the $O\left(
p_{x}^{4}\right)  $ term from the lattice dispersion relation, this time we
tune the coefficient of the $O\left(  p_{x}^{4}\right)  $ term to match as
best as possible the continuum dispersion relation $\omega(p_{x})=p_{x}%
^{2}/(2m)$ over the full range $-\Lambda\leq p_{x}\leq\Lambda$. \ Let us
define the $O(a^{2})$-well-tempered\ lattice kinetic energy action
$K^{\text{(wt1)}}$ and dispersion relation $\omega^{\text{(wt1)}}(p_{x})$,%
\begin{align}
K^{\text{(wt1)}}  &  =K^{(0)}+s\left(  K^{(1)}-K^{(0)}\right)  ,\\
\omega^{\text{(wt1)}}(p_{x})  &  =\omega^{(0)}(p_{x})+s\left(  \omega
^{(1)}(p_{x})-\omega^{(0)}(p_{x})\right)  , \label{disprelwt}%
\end{align}
where the unknown coefficient $s$ is given by the integral constraint%
\begin{equation}%
{\displaystyle\int\limits_{-\Lambda}^{\Lambda}}
dp_{x}\;\omega^{\text{(wt1)}}(p_{x})=%
{\displaystyle\int\limits_{-\Lambda}^{\Lambda}}
dp_{x}\;\frac{p_{x}^{2}}{2m}.
\end{equation}
Solving for $s$ gives $s=\frac{2}{3}\pi^{2}-4\approx2.5797$. \ We show a
comparison of the lattice dispersion relations $\omega^{(0)}$, $\omega^{(1)}$,
$\omega^{\text{(wt1)}}$, and the continuum limit $\omega$ in Fig.
\ref{dispersion2}.
Matching $E_{0}(p_{x}=0)=-2.348$\ MeV for the well-tempered lattice action
gives an adjusted coefficient of $C(\Lambda)=-0.1260$. \ Results for the
two-particle ground state and lowest scattering state for the well-tempered
action are shown in Fig. \ref{welltempered_1d}.
The deviation between lattice and continuum results for the ground state
energy has been substantially reduced.

\section{Zero-range neutrons in three dimensions}

We now explore how various lattice actions affect cutoff errors in three
dimensions. \ We consider spin-1/2 fermions in three dimensions with
zero-range attraction. \ This gives an approximate description of interacting
neutrons below $1\%$ of nuclear matter density. \ To demonstrate the
generality of our lattice error analysis we consider both the grand canonical
ensemble in the Euclidean lattice formalism as well as the canonical ensemble
in the Hamiltonian lattice formalism.

In the continuum limit the Hamiltonian for zero-range neutrons has the form%
\begin{equation}
H=-\frac{1}{2m}\sum_{i=\uparrow,\downarrow}\int d^{3}\vec{r}\;a_{i}^{\dagger
}(\vec{r})\vec{\nabla}^{2}a_{i}(\vec{r})+C\int d^{3}\vec{r}\;a_{\downarrow
}^{\dagger}(\vec{r})a_{\uparrow}^{\dagger}(\vec{r})a_{\uparrow}(\vec
{r})a_{\downarrow}(\vec{r}). \label{continuum_3D}%
\end{equation}
Just as in our one-dimensional model, the connected amputated two-particle
Green's function for zero-range neutrons in three dimensions is given by the
sum of chained bubble diagrams shown in Fig. \ref{twotwo}. \ Any connected
scattering process can be constructed from two-particle Green's functions
linked together with free particle propagators. \ While the two-particle
Green's function is divergent, all of the new loop integrations produced by
connecting two-neutron Green's functions are ultraviolet finite.

Let $G_{2}(p_{0},\vec{p})$ be the amplitude for the connected amputated
two-particle Green's function, where$\ p_{0}$ is the total energy and $\vec
{p}$ is the total spatial momentum of the two particles. \ We sum the bubble
diagrams in Fig. \ref{twotwo} and get%
\begin{equation}
G_{2}(p_{0},\vec{p})=\frac{-iC}{1-iC\cdot\Pi\left(  p_{0},\vec{p}\right)
}=\frac{1}{-\frac{1}{iC}+\Pi\left(  p_{0},\vec{p}\right)  },
\end{equation}
where
\begin{equation}
\Pi\left(  p_{0},\vec{p}\right)  =\int\frac{dq_{0}d^{3}\vec{q}}{\left(
2\pi\right)  ^{4}}\frac{i}{\frac{p_{0}}{2}+q_{0}-\frac{\left(  \frac{\vec{p}%
}{2}+\vec{q}\right)  ^{2}}{2m}+i\varepsilon}\times\frac{i}{\frac{p_{0}}%
{2}-q_{0}-\frac{\left(  \frac{\vec{p}}{2}-\vec{q}\right)  ^{2}}{2m}%
+i\varepsilon}.
\end{equation}
Since $\Pi\left(  p_{0},\vec{p}\right)  $ is ultraviolet divergent we
renormalize the coupling $C$ to absorb the divergence. \ In the end we get%
\begin{equation}
G_{2}(p_{0},\vec{p},\Lambda)=\frac{i4\pi/m}{-\frac{1}{a_{\text{scatt}}}%
-i\sqrt{mp_{0}-\frac{\vec{p}^{2}}{4}}+\Lambda\cdot O\left(  \frac{Q^{2}%
}{\Lambda^{2}}\right)  }, \label{3DGreenFn}%
\end{equation}
where $a_{\text{scatt}}$ is the s-wave scattering length and $Q^{2}$ is some
homogeneous combination of the parameters $mp_{0}$ and $\vec{p}^{2}$ which
depends on the regularization scheme. \ The cutoff error can be regarded as a
momentum/energy-dependent $O(Q^{2}/\Lambda)$ modification to the inverse
scattering length.

\subsection{One- and two-particle lattice dispersion relation in three
dimensions}

Just as in the one-dimensional case we start with the simplest possible
lattice Hamiltonian that reproduces (\ref{continuum_3D}) in the continuum
limit. \ We let%
\begin{equation}
H^{(0)}=K^{(0)}+V^{(0)},
\end{equation}%
\begin{equation}
K^{(0)}=\frac{3}{m}\sum_{\vec{n}_{s},i}a_{i}^{\dagger}(\vec{n}_{s})a_{i}%
(\vec{n}_{s})-\frac{1}{2m}\sum_{\vec{n}_{s},\hat{l}_{s},i}\left[
a_{i}^{\dagger}(\vec{n}_{s})a_{i}(\vec{n}_{s}+\hat{l}_{s})+a_{i}^{\dagger
}(\vec{n}_{s})a_{i}(\vec{n}_{s}-\hat{l}_{s})\right]  , \label{standardK_3d}%
\end{equation}%
\begin{equation}
V^{(0)}=C\sum_{\vec{n}_{s}}a_{\downarrow}^{\dagger}(\vec{n}_{s})a_{\uparrow
}^{\dagger}(\vec{n}_{s})a_{\uparrow}(\vec{n}_{s})a_{\downarrow}(\vec{n}_{s}).
\end{equation}
Here $\vec{n}_{s}$ is a three-dimensional spatial lattice vector and $\hat
{l}_{s}=\hat{x},\hat{y},\hat{z}$ are lattice unit vectors in each of the 3
spatial directions. \ The $s$ subscript is our notation for spatial lattice
vectors with no time component. $\ H^{(0)}$ is the standard lattice
Hamiltonian. \ The zero superscript again signifies that it is the simplest
possible lattice formulation. \ We take the same values $m=939$ MeV for the
neutron mass and $a=(50$ MeV$)^{-1}$ for the lattice spacing. \ This again
yields $\Lambda=\pi a^{-1}\approx157$ MeV for the cutoff momentum. \ The
single-particle dispersion relation for the standard lattice Hamiltonian is%
\begin{equation}
\omega^{(0)}(\vec{p}_{s})=\frac{1}{m}\sum_{l_{s}=x,y,z}\left(  1-\cos
p_{l_{s}}\right)  =\frac{\vec{p}_{s}^{2}}{2m}+O\left(  \left\vert \vec{p}%
_{s}\right\vert ^{4}\right)  .
\end{equation}

We consider a lattice system which is a periodic cubic lattice of length $L$.
\ If we set the two-body scattering pole in the rest frame at energy
$E_{\text{pole}}$ then the cutoff-dependent coefficient $C(\Lambda)$ satisfies%
\begin{equation}
-\frac{1}{C(\Lambda)}=\lim_{L\rightarrow\infty}\frac{1}{L^{3}}\sum_{\vec
{k}_{s}}\frac{1}{-E_{\text{pole}}+2\omega^{(0)}(2\pi\vec{k}_{s}/L)},
\end{equation}
where the components of $\vec{k}_{s}$ are integers from $0,1,\cdots,L-1$. \ If
there is a two-body bound state then we can take $E_{\text{pole}}$ equal to
negative the binding energy. \ Alternatively we can choose $E_{\text{pole}}$
to be the pole nearest threshold and use L\"{u}scher's formula for the finite
volume two-body spectrum \cite{Luscher:1986pf,Beane:2003da},%
\begin{equation}
E_{0}(L)=\dfrac{4\pi a_{\text{scatt}}}{m_{N}L^{3}}[1-c_{1}\frac
{a_{\text{scatt}}}{L}+c_{2}\frac{a_{\text{scatt}}^{2}}{L^{2}}+\cdots],
\end{equation}
where $c_{1}=-2.837297,$ $c_{2}=6.375183.$

\begin{figure}
[ptb]
\begin{center}
\includegraphics[
height=2.5244in,
width=3.5276in
]%
{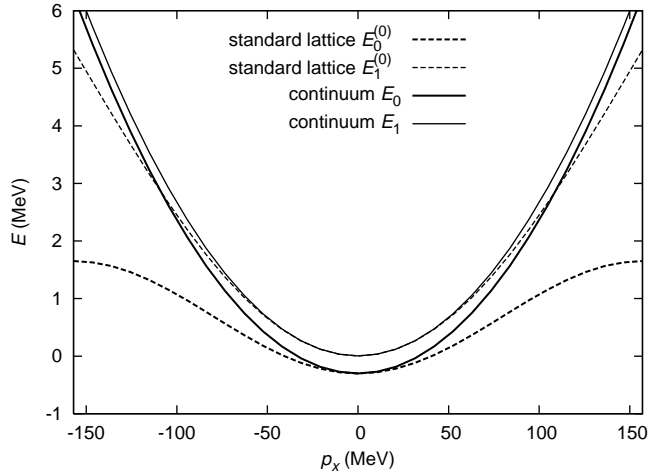}%
\caption{The lowest two-particle energies for the standard lattice action,
$E_{0}^{(0)}$ and $E_{1}^{(0)}$, and the corresponding continuum limit values,
$E_{0}$ and $E_{1}$.}%
\label{standard_3d}%
\end{center}
\end{figure}

As an example we set $E_{\text{pole}}=-0.300$ MeV and find that $C(\Lambda
)=-9.464\times10^{-5}$ MeV$^{-2}$. \ This corresponds with a scattering length
$a_{\text{scatt}}=11.76$ fm. \ For $L=30$ we plot the energy of the lowest two
energy states $E_{0}^{(0)}$ and $E_{1}^{(0)}$ as a function of the total
momentum in Fig. \ref{standard_3d} and compare with the corresponding
continuum limit values, $E_{0}$ and $E_{1}$. \ In physical units $L=30$
corresponds with $118$ fm. \ This is about ten times the scattering length and
so finite volume effects are negligible. \ The box length does however
determine the level spacing between unbound scattering states. \ We take the
total momentum $\vec{p}_{s}$ along the $x$-axis so that $\vec{p}_{s}%
=(p_{x},0,0)$. \
Just as we found in the one-dimensional model at strong coupling, we encounter
the same problem of broken Galilean invariance. \ While the agreement between
the excited states $E_{1}^{(0)}$ and $E_{1}$ is not bad, the deviation between
lattice and continuum results for the ground state energy is substantial for
$p_{x}\neq0$. \ Since we have chosen the lattice coupling $C(\Lambda)$ so that
$E_{0}^{(0)}=E_{0}$ at $p_{x}=0,$ the disagreement between $E_{0}^{(0)}$ and
$E_{0}$ is proportional $p_{x}^{2}.$

\subsection{$O(a^{2})$-improved and $O(a^{2})$-well-tempered actions in three
dimensions}

\begin{figure}
[ptb]
\begin{center}
\includegraphics[
height=2.5244in,
width=3.5276in
]%
{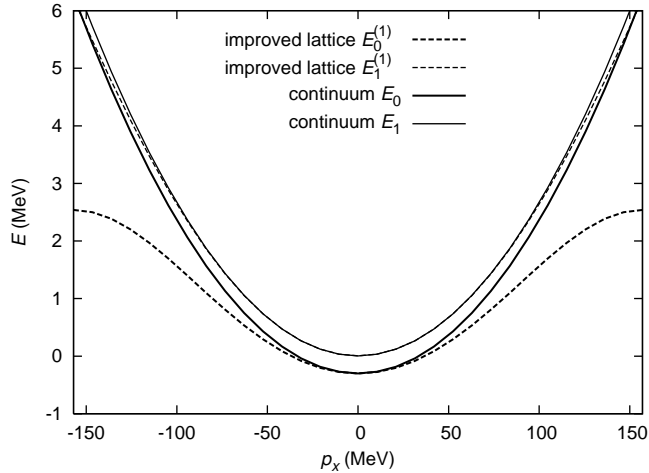}%
\caption{The lowest two-particle energies for the improved lattice action,
$E_{0}^{(1)}$ and $E_{1}^{(1)}$, and the continuum energies, $E_{0}$ and
$E_{1}$.}%
\label{improved_3d}%
\end{center}
\end{figure}

Just as in the one-dimensional case we can replace the standard lattice
kinetic energy with an $O(a^{2})$-improved kinetic energy,%
\begin{align}
K^{(1)}  &  =\frac{5}{4}\times\frac{3}{m}\sum_{\vec{n}_{s},i}a_{i}^{\dagger
}(\vec{n}_{s})a_{i}(\vec{n}_{s})-\frac{4}{3}\times\frac{1}{2m}\sum_{\vec
{n}_{s},\hat{l}_{s},i}\left[  a_{i}^{\dagger}(\vec{n}_{s})a_{i}(\vec{n}%
_{s}+\hat{l}_{s})+a_{i}^{\dagger}(\vec{n}_{s})a_{i}(\vec{n}_{s}-\hat{l}%
_{s})\right] \nonumber\\
&  +\frac{1}{12}\times\frac{1}{2m}\sum_{\vec{n}_{s},\hat{l}_{s},i}\left[
a_{i}^{\dagger}(\vec{n}_{s})a_{i}(\vec{n}_{s}+2\hat{l}_{s})+a_{i}^{\dagger
}(\vec{n}_{s})a_{i}(\vec{n}_{s}-2\hat{l}_{s})\right]  .
\end{align}
This gives the dispersion relation%
\begin{equation}
\omega^{(1)}(\vec{p}_{s})=\frac{1}{m}\sum_{l_{s}=x,y,z}\left[  \frac{5}%
{4}-\frac{4}{3}\cos p_{l_{s}}+\frac{1}{12}\cos\left(  2p_{l_{s}}\right)
\right]  =\frac{\vec{p}_{s}^{2}}{2m}+O\left(  \left\vert \vec{p}%
_{s}\right\vert ^{6}\right)  . \label{omega1}%
\end{equation}
In this case the renormalization condition for $C(\Lambda)$ is%
\begin{equation}
-\frac{1}{C(\Lambda)}=\lim_{L\rightarrow\infty}\frac{1}{L^{3}}\sum_{\vec
{k}_{s}}\frac{1}{-E_{\text{pole}}+2\omega^{(1)}(2\pi\vec{k}_{s}/L)},
\end{equation}
and for $E_{\text{pole}}=-0.300$ MeV we find $C(\Lambda)=-1.1031\times10^{-4}$
MeV$^{-2}$. \ Results for the two-particle ground state and lowest scattering
state with the improved action are shown in Fig. \ref{improved_3d}.%
The results are somewhat better for the $O(a^{2})$-improved kinetic energy,
though the agreement between $E_{1}^{(1)}$ and $E_{1}$ all the way up to the
cutoff momentum should be regarded as accidental. \ As in the one-dimensional
case, we expect that better agreement may be possible for the ground state
using a well-tempered action.

\begin{figure}
[ptb]
\begin{center}
\includegraphics[
height=2.5244in,
width=3.5276in
]%
{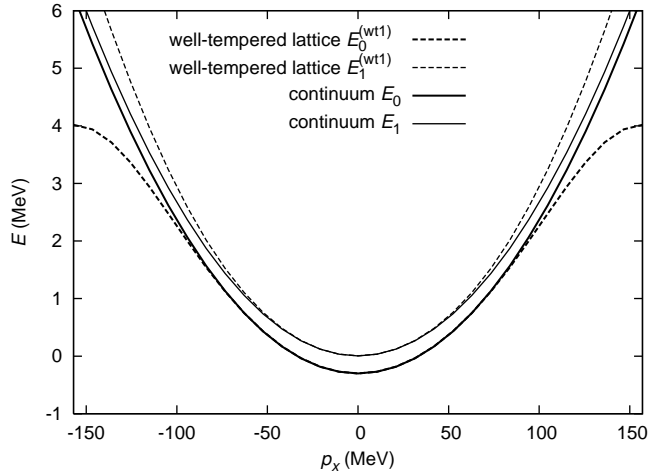}%
\caption{The lowest two-particle energies for the well-tempered lattice
action, $E_{0}^{(\text{wt}1)}$ and $E_{1}^{(\text{wt}1)}$, and the continuum
energies, $E_{0}$ and $E_{1}$.}%
\label{welltempered_3d}%
\end{center}
\end{figure}

We define the $O(a^{2})$-well-tempered kinetic energy as%
\begin{equation}
K^{\text{(wt1)}}=K^{(0)}+s\left(  K^{(1)}-K^{(0)}\right)  ,
\end{equation}
where $s$ is given by the following integral constraint on the resulting
dispersion relation:%
\begin{equation}%
{\displaystyle\int\limits_{-\Lambda}^{\Lambda}}
{\displaystyle\int\limits_{-\Lambda}^{\Lambda}}
{\displaystyle\int\limits_{-\Lambda}^{\Lambda}}
dp_{x}dp_{y}dp_{z}\;\omega^{\text{(wt1)}}(\vec{p}_{s})=%
{\displaystyle\int\limits_{-\Lambda}^{\Lambda}}
{\displaystyle\int\limits_{-\Lambda}^{\Lambda}}
{\displaystyle\int\limits_{-\Lambda}^{\Lambda}}
dp_{x}dp_{y}dp_{z}\;\frac{\vec{p}_{s}^{2}}{2m}.
\end{equation}
Since both $\omega^{\text{(wt1)}}(\vec{p}_{s})$ and $\vec{p}_{s}^{2}/(2m)$
decompose as a sum of separate terms for $p_{x}$, $p_{y}$, and $p_{z}$, this
gives the same result as in the one-dimensional case, $s=\frac{2}{3}\pi
^{2}-4\approx2.5797$. \ For the well-tempered action
\begin{equation}
-\frac{1}{C(\Lambda)}=\lim_{L\rightarrow\infty}\frac{1}{L^{3}}\sum_{\vec
{k}_{s}\text{ integer}}\frac{1}{-E_{\text{pole}}+2\omega^{(\text{wt}1)}%
(2\pi\vec{k}_{s}/L)},
\end{equation}
and for $E_{\text{pole}}=-0.300$ MeV we find $C(\Lambda)=-1.3273\times10^{-4}$
MeV$^{-2}$. \ Results for the two-particle ground state and lowest scattering
state with the well-tempered action are shown in Fig. \ref{welltempered_3d}.
Just as in the one-dimensional model, we find the deviation between lattice
and continuum results for the ground state energy has been substantially reduced.

The well-tempered kinetic energy appears to fix much of the strongly-broken
Galilean invariance on the lattice. \ In the remainder of our analysis we see
if it also fixes the large discretization errors at nonzero temperature. \ We
do this by calculating the second virial coefficient $b_{2}(T)$, which was
found to have large discretization errors in \cite{Lee:2005is}. \ There are
several ways to calculate $b_{2}(T)$ on the lattice, and it is not obvious
that the lattice errors are the same for different calculations. \ Therefore
in the next two sections we consider two different lattice calculations of
$b_{2}(T)$. \ The first calculation relies on the virial expansion of the
density in the grand canonical ensemble,%
\begin{equation}
\rho=\frac{2}{\lambda_{T}^{3}}\left[  z+2b_{2}(T)z^{2}+\cdots\right]  .
\end{equation}
\ We use the Euclidean lattice formalism with nonzero temporal lattice spacing
for this calculation. \ The second method finds $b_{2}(T)$ by means of the
two-particle partition function,%
\begin{equation}
b_{2}(T)-b_{2}^{\text{free}}(T)=\frac{\lambda_{T}^{3}}{2V}\left\{
Tr_{2}[\text{exp}(-\beta H)]-Tr_{2}[\text{exp}(-\beta H_{\text{free}%
})]\right\}  .
\end{equation}
We use the Hamiltonian lattice formalism for this calculation.

\section{Grand canonical Euclidean lattice calculation for $b_{2}(T)$}

In this section we calculate the second virial coefficient $b_{2}(T)$ in the
grand canonical ensemble using the Euclidean lattice formalism. \ We use the
same values $m=939$ MeV for the neutron mass and $a=(50$ MeV$)^{-1}$ for the
lattice spacing. \ We set the temporal lattice spacing at $a_{t}=(24$
MeV$)^{-1}$. \ These are the same values as used in
\cite{Lee:2005is,Lee:2005it}. \ We define $\alpha_{t}$ as the ratio of
temporal to spatial lattice spacings. \ In our notation $\vec{n}=(n_{t}%
,\vec{n}_{s})$ denotes space-time lattice vectors. $\ c$ and $c^{\ast}$ are
Grassmann variables for the neutrons in the path integral formalism.
$\ \hat{0}$ is a lattice unit vector in the temporal direction. \ $\hat{l}%
_{s}=\hat{x},\hat{y},\hat{z}$ are lattice unit vectors for the spatial
directions. $\ $Also $\mu$ is the chemical potential, $L$ is the spatial
length of the cubic lattice, and $L_{t}$ is the temporal length.

In the grand canonical ensemble the partition function can be written as%
\begin{equation}
\mathcal{Z}\propto\int DcDc^{\ast}\exp\left(  -S\right)  ,
\end{equation}
where%
\begin{equation}
S=S_{\text{free}}+C^{\prime}\alpha_{t}e^{2\mu\alpha_{t}}\sum_{\vec{n}%
}c_{\downarrow}^{\ast}(\vec{n})c_{\uparrow}^{\ast}(\vec{n})c_{\uparrow}%
(\vec{n})c_{\downarrow}(\vec{n}),
\end{equation}
and the standard free lattice action is given by%
\begin{align}
S_{\text{free}}^{(0)}  &  =\sum_{\vec{n},i}\left[  c_{i}^{\ast}(\vec{n}%
)c_{i}(\vec{n}+\hat{0})-e^{\mu\alpha_{t}}c_{i}^{\ast}(\vec{n})c_{i}(\vec
{n})\right]  +\alpha_{t}e^{\mu\alpha_{t}}\times\frac{3}{m}\sum_{\vec{n}%
,i}c_{i}^{\ast}(\vec{n})c_{i}(\vec{n})\nonumber\\
&  -\alpha_{t}e^{\mu\alpha_{t}}\times\frac{1}{2m}\sum_{\vec{n},\hat{l}_{s}%
,i}\left[  c_{i}^{\ast}(\vec{n})c_{i}(\vec{n}+\hat{l}_{s})+c_{i}^{\ast}%
(\vec{n})c_{i}(\vec{n}-\hat{l}_{s})\right]  .
\end{align}
Our coupling constant $C^{\prime}$ differs from the coupling constant $C$
appearing in \cite{Lee:2004qd,Lee:2005is,Lee:2005it}. \ However the two are
simply related,%
\begin{equation}
C^{\prime}\alpha_{t}=-\left(  e^{-C\alpha_{t}}-1\right)  \left(
1-\frac{3\alpha_{t}}{m}\right)  ^{2}.
\end{equation}

Let us define%
\begin{equation}
S_{\text{static}}=\sum_{\vec{n},i}\left[  c_{i}^{\ast}(\vec{n})c_{i}(\vec
{n}+\hat{0})-e^{\mu\alpha_{t}}c_{i}^{\ast}(\vec{n})c_{i}(\vec{n})\right]  .
\end{equation}
Then we have%
\begin{equation}
S_{\text{free}}^{(0)}=S_{\text{static}}+\alpha_{t}e^{\mu\alpha_{t}%
}S_{\text{kinetic}}^{(0)},
\end{equation}
where%
\begin{equation}
S_{\text{kinetic}}^{(0)}=\frac{3}{m}\sum_{\vec{n},i}c_{i}^{\ast}(\vec{n}%
)c_{i}(\vec{n})-\frac{1}{2m}\sum_{\vec{n},\hat{l}_{s},i}\left[  c_{i}^{\ast
}(\vec{n})c_{i}(\vec{n}+\hat{l}_{s})+c_{i}^{\ast}(\vec{n})c_{i}(\vec{n}%
-\hat{l}_{s})\right]  .
\end{equation}
$S_{\text{kinetic}}^{(0)}$ is analogous with $K^{(0)}$ in (\ref{standardK_3d}%
). \ The $O(a^{2}$)-improved action has the form%
\[
S_{\text{free}}^{(1)}=S_{\text{static}}+\alpha_{t}e^{\mu\alpha_{t}%
}S_{\text{kinetic}}^{(1)},
\]
where%
\begin{align}
S_{\text{kinetic}}^{(1)}  &  =\frac{5}{4}\times\frac{3}{m}\sum_{\vec{n}%
,i}c_{i}^{\ast}(\vec{n})c_{i}(\vec{n})-\frac{4}{3}\times\frac{1}{2m}\sum
_{\vec{n},\hat{l}_{s},i}\left[  c_{i}^{\ast}(\vec{n})c_{i}(\vec{n}+\hat{l}%
_{s})+c_{i}^{\ast}(\vec{n})c_{i}(\vec{n}-\hat{l}_{s})\right] \nonumber\\
&  +\frac{1}{12}\times\frac{1}{2m}\sum_{\vec{n}_{s},\hat{l}_{s},i}\left[
c_{i}^{\ast}(\vec{n})c_{i}(\vec{n}+2\hat{l}_{s})+c_{i}^{\ast}(\vec{n}%
)c_{i}(\vec{n}-2\hat{l}_{s})\right]  .
\end{align}
The $O(a^{2}$)-well-tempered action has the form%
\begin{equation}
S_{\text{free}}^{(\text{wt}1)}=S_{\text{static}}+\alpha_{t}e^{\mu\alpha_{t}%
}S_{\text{kinetic}}^{(\text{wt}1)}%
\end{equation}
where%
\begin{equation}
S_{\text{kinetic}}^{(\text{wt}1)}=S_{\text{kinetic}}^{(\text{0})}+s\left(
S_{\text{kinetic}}^{(1)}-S_{\text{kinetic}}^{(0)}\right)  ,
\end{equation}%
\begin{equation}
s=\frac{2}{3}\pi^{2}-4\approx2.5797.
\end{equation}

For each lattice action we define the free neutron propagator,%
\begin{equation}
D_{\text{free}}(\vec{k})=\frac{\int DcDc^{\ast}\;\tilde{c}_{i}(\vec{k}%
)\tilde{c}_{i}^{\ast}(-\vec{k})\exp\left(  -S_{\text{free}}\right)  }{\int
DcDc^{\ast}\;\exp\left(  -S_{\text{free}}\right)  }\text{ (no sum on
}i\text{),}%
\end{equation}
where the components of $\vec{k}=(k_{0},\vec{k}_{s})$ are integers. \ Our
conventions for the lattice Fourier transform are%
\begin{align}
\tilde{f}(\vec{k})  &  =\sum_{\vec{n}}e^{i\frac{2\pi n_{t}k_{0}}{L_{t}}%
}e^{i\frac{2\pi\vec{n}_{s}\cdot\vec{k}_{s}}{L}}f(\vec{n}),\\
f(\vec{n})  &  =\frac{1}{L_{t}L^{3}}\sum_{\vec{k}}e^{-i\frac{2\pi n_{t}k_{0}%
}{L_{t}}}e^{-i\frac{2\pi\vec{n}_{s}\cdot\vec{k}_{s}}{L}}\tilde{f}(\vec{k}).
\end{align}
Let $\omega(2\pi\vec{k}_{s}/L)$ be the lattice dispersion relation, either
$\omega^{(0)}(2\pi\vec{k}_{s}/L)$, $\omega^{(1)}(2\pi\vec{k}_{s}/L),$ or
$\omega^{(\text{wt1)}}(2\pi\vec{k}_{s}/L)$ as defined in the previous section.
\ Then we have%
\begin{equation}
D_{\text{free}}(\vec{k})=\frac{1}{e^{-i\frac{2\pi k_{0}}{L_{t}}}-e^{\mu
\alpha_{t}}+\alpha_{t}e^{\mu\alpha_{t}}\omega(2\pi\vec{k}_{s}/L)}.
\end{equation}
In \cite{Lee:2004qd} it was shown that the cutoff-dependent coupling constant
is given by the constraint%
\begin{equation}
-\frac{1}{\alpha_{t}C^{\prime}(\Lambda)}=\lim_{L\rightarrow\infty}\frac
{1}{L^{3}}\sum_{\vec{k}_{s}}\frac{1}{e^{-\alpha_{t}E_{\text{pole}}}-\left[
1-\alpha_{t}\omega(2\pi\vec{k}_{s}/L)\right]  ^{2}}.
\end{equation}

\begin{figure}
[ptb]
\begin{center}
\includegraphics[
height=1.4615in,
width=2.847in
]%
{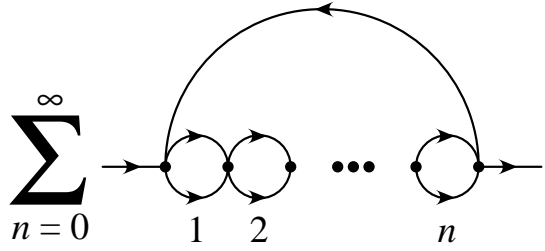}%
\caption{Two-particle bubble diagrams contributing to the neutron self-energy
to order $z^{2}$.}%
\label{selfenergy}%
\end{center}
\end{figure}

We use the Euclidean lattice action to compute the neutron density as a
function of temperature, chemical potential, and interaction strength. \ Let
$\rho_{\text{free}}$ be the free neutron density and $\rho$ be the neutron
density with interactions. \ Then from the virial expansion we get%
\begin{equation}
\rho-\rho_{\text{free}}=\frac{4}{\lambda_{T}^{3}}\left[  b_{2}(T)-b_{2}%
^{\text{free}}(T)\right]  z^{2}+O(z^{3}).
\end{equation}
We note that our convention for the second lattice coefficient $b_{2}(T)$ is
slightly different from the one used in \cite{Lee:2005is}. \ The densities
$\rho_{\text{free}}$ and $\rho$ are computed using the free and full neutron
propagators,%
\begin{align}
\rho_{\text{free}}  &  =\frac{1}{\beta L^{3}}\frac{\partial}{\partial\mu}%
\ln\mathcal{Z}_{\text{free}}=2\left[  1-\frac{1}{L_{t}L^{3}}\sum_{\vec{k}%
}D_{\text{free}}(\vec{k})e^{-i\frac{2\pi k_{0}}{L_{t}}}\right]  ,\\
\rho &  =\frac{1}{\beta L^{3}}\frac{\partial}{\partial\mu}\ln\mathcal{Z}%
=2\left[  1-\frac{1}{L_{t}L^{3}}\sum_{\vec{k}}D(\vec{k})e^{-i\frac{2\pi k_{0}%
}{L_{t}}}\right]  .
\end{align}
The full neutron propagator $D(\vec{k})$ can be expressed in terms of the
neutron self-energy, $\Sigma(\vec{k})$,
\begin{equation}
D(\vec{k})=\frac{D_{\text{free}}(\vec{k})}{1-\Sigma(\vec{k})D_{\text{free}%
}(\vec{k})}. \label{Dfull}%
\end{equation}
We compute the self-energy to order $z^{2}$ by summing the two-particle bubble
diagrams shown in Fig. \ref{selfenergy}. \
Further details of the calculation can be found in \cite{Lee:2004qd}.

In addition to these local actions we also consider a dispersion relation
given by%
\begin{equation}
\omega^{(\text{quad})}(2\pi\vec{k}_{s}/L)=\frac{1}{2m}\sum_{l_{s}%
=x,y,z}\left(  \frac{2\pi}{L}k_{l_{s}}^{\prime}\right)  ^{2},
\end{equation}
where%
\begin{equation}
k_{l_{s}}^{\prime}\equiv\operatorname{mod}(k_{l_{s}},L),\text{\quad}\left\vert
k_{l_{s}}^{\prime}\right\vert \leq L/2.
\end{equation}
This quadratic dispersion relation was used in \cite{Bulgac:2005a} to reduce
cutoff effects. \ Since it equals the continuum dispersion relation for
$\left\vert 2\pi\vec{k}_{s}/L\right\vert \leq\Lambda$ we expect it also to be
effective in reducing errors due to broken Galilean invariance. \ However it
does not correspond with a local lattice action. \ It was implemented in
\cite{Bulgac:2005a} by Fourier transforming back and forth between position
space and momentum space. \ Unfortunately this results in a steeper
computational scaling for the Monte Carlo algorithm as a function of volume.
\ Nevertheless there is no significant computational problem for the
perturbative calculation presented here, and so we include it in our analysis
for comparison.

\begin{figure}
[ptb]
\begin{center}
\includegraphics[
height=4.5222in,
width=3.1756in,
angle=-90
]%
{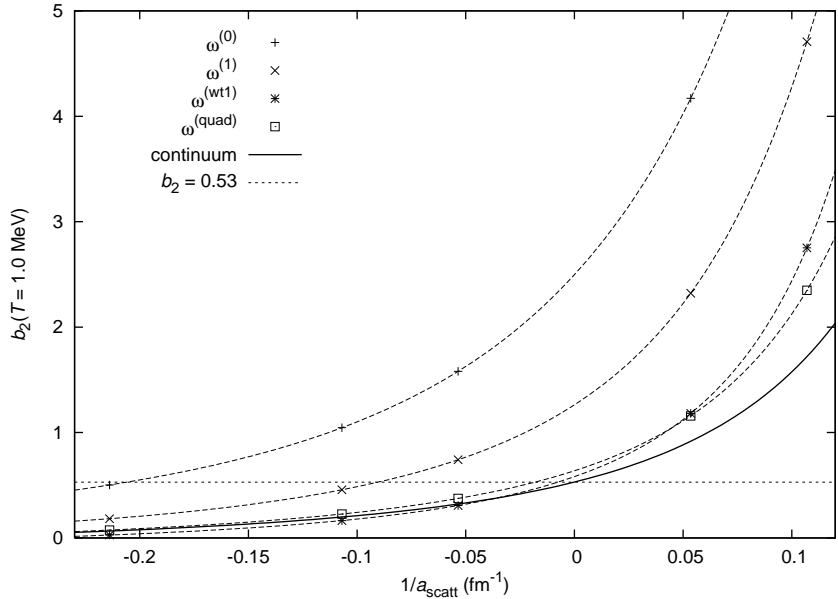}%
\caption{Plot of $b_{2}(T)$ for $\omega^{(0)}$, $\omega^{(1)}$, $\omega
^{(\text{wt1})}$, $\omega^{(\text{quad})}$ and the continuum limit for $T$ =
1.0 MeV.}%
\label{b2t100}%
\end{center}
\end{figure}

\begin{figure}
[ptbptb]
\begin{center}
\includegraphics[
height=4.5221in,
width=3.1756in,
angle=-90
]%
{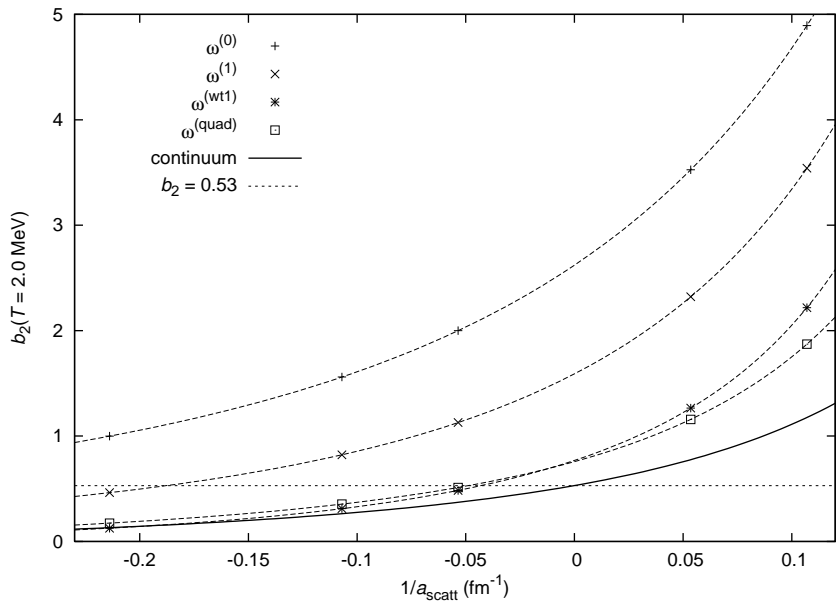}%
\caption{Plot of $b_{2}(T)$ for $\omega^{(0)}$, $\omega^{(1)}$, $\omega
^{(\text{wt1})}$, $\omega^{(\text{quad})}$ and the continuum limit for $T$ =
2.0 MeV.}%
\label{b2t200}%
\end{center}
\end{figure}

We can compute $b_{2}(T)$ at any small fugacity, and so we choose
$z=e^{-5}\approx0.0067$. \ We take the lattice length to be $L=8$, which is
sufficiently large enough that the finite volume error for the local actions
is less than $1\%$. \ The non-local action associated with $\omega
^{(\text{quad})}$ appears to have a slightly larger finite volume error.
\ Using each of these dispersion relations, we compute the second virial
coefficient for a range of scattering lengths, $a_{\text{scatt}}=-4.675,$
$-9.35$, $-18.70$, $+18.70$, $+9.35$ fm$.$ \ In Fig. \ref{b2t100} we show the
results for $b_{2}(T)$ as a function of inverse scattering length for
dispersion relations $\omega^{(0)}$, $\omega^{(1)}$, $\omega^{(\text{wt1})}$,
and $\omega^{(\text{quad})}$ at $T$ = 1.0 MeV. \ We also show the continuum
limit result given in (\ref{b2}). \ Analogous results at temperature $T$ = 2.0
MeV are shown in Fig. \ref{b2t200}.%
We see that of the various lattice dispersion relations, $\omega
^{(\text{wt1})}$ and $\omega^{(\text{quad})}$ come closest to the continuum limit.

We can compare the different lattice actions in a slightly different way.
\ Let us think of the temperature as fixed and the scattering length as
varying. \ When deriving (\ref{3DGreenFn}) we found that the finite cutoff
error can be regarded as a momentum/energy-dependent $O(Q^{2}/\Lambda)$
modification to the inverse scattering length. \ In the continuum limit
$b_{2}(T)=3\times2^{-\frac{5}{2}}$ at infinite scattering length for all $T$.
\ At finite cutoff let $a_{\text{scatt}}^{\infty}(T)$ be the scattering length
for which $b_{2}(T)=3\times2^{-\frac{5}{2}}\approx0.530$. \ We can now
interpret the cutoff error as a small modification to the inverse scattering
length, $\Delta a_{\text{scatt}}^{-1}(T)=-1/a_{\text{scatt}}^{\infty}$. \ In
the continuum limit $\Delta a_{\text{scatt}}^{-1}(T)=0$ for all $T$, and the
shift provides a simple quantitative measure of the cutoff error near infinite
scattering length.

\begin{figure}
[ptb]
\begin{center}
\includegraphics[
height=4.5703in,
width=3.2088in,
angle=-90
]%
{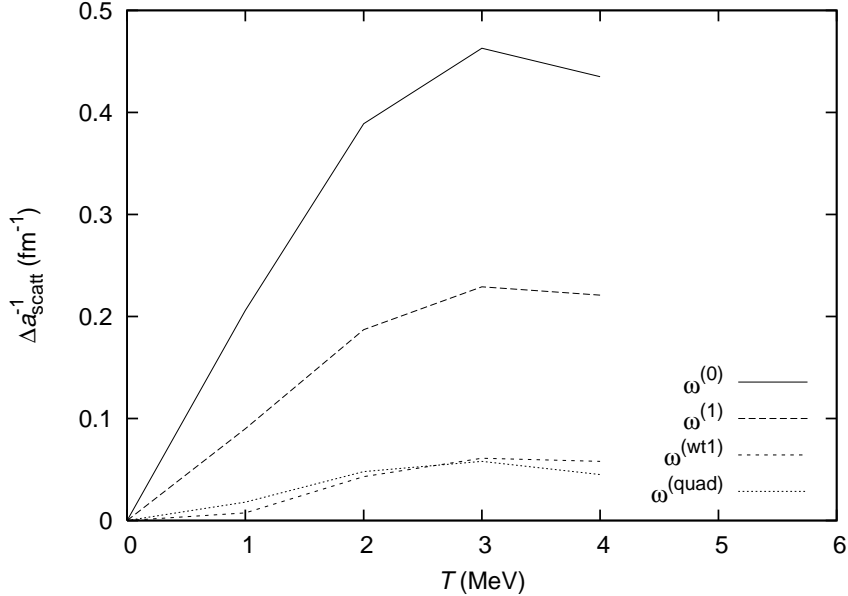}%
\caption{Plot of $\Delta a_{\text{scatt}}^{-1}$ as a function of temperature
for the dispersion relations $\omega^{(0)}$, $\omega^{(1)}$, $\omega
^{(\text{wt1})}$, and $\omega^{(\text{quad})}$.}%
\label{delta}%
\end{center}
\end{figure}

We expect broken Galilean invariance due to the cutoff to introduce a term of
size $O(\vec{p}_{s}^{2}/\Lambda)$ in $\Delta a_{\text{scatt}}^{-1}(T)$. \ In
the classical regime we know from the equipartition theorem that the average
value of $\vec{p}_{s}^{2}$ scales linearly with the temperature $T$.
\ Therefore we expect $\Delta a_{\text{scatt}}^{-1}(T)$ also to scale linearly
with $T$. \ In Fig. \ref{delta} we plot $\Delta a_{\text{scatt}}^{-1}(T)$ for
the dispersion relations $\omega^{(0)}$, $\omega^{(1)}$, $\omega
^{(\text{wt1})}$, and $\omega^{(\text{quad})}$. \
We see the expected linear dependence in $\Delta a_{\text{scatt}}^{-1}(T)$ for
small $T$. \ We also see that $\Delta a_{\text{scatt}}^{-1}(T)$ for
$\omega^{(\text{wt1})}$ and $\omega^{(\text{quad})}$ are quite a bit smaller
that $\Delta a_{\text{scatt}}^{-1}(T)$ for $\omega^{(\text{0})}$ and
$\omega^{(\text{1})}$. \ In fact most of the cutoff error at nonzero $T$
appears to have been removed.

\section{Two-particle Hamiltonian lattice calculation for $b_{2}(T)$}

\begin{figure}
[ptb]
\begin{center}
\includegraphics[
height=4.5697in,
width=3.2085in,
angle=-90
]%
{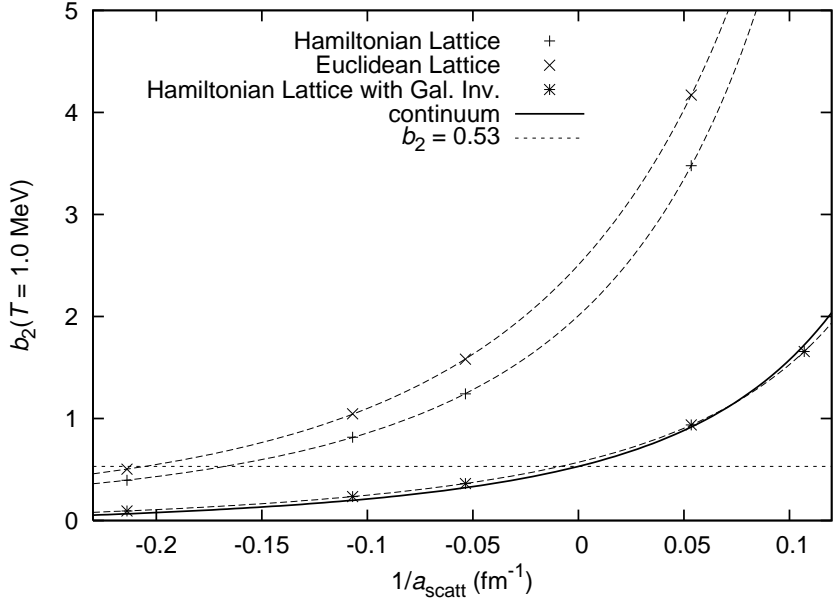}%
\caption{Plot of $b_{2}$ at $T$ = 1.0 MeV for the standard Hamiltonian lattice
action at $T=1$.$0$ MeV. \ For comparison we show results for the standard
Euclidean lattice action, the standard Hamiltonian lattice action with
Galilean invariance imposed by hand, and the continuum limit.}%
\label{b2_hop1}%
\end{center}
\end{figure}

In this section we return to the Hamiltonian lattice formalism and compute
$b_{2}(T)$ using the two-particle trace formula,%
\begin{equation}
b_{2}(T)-b_{2}^{\text{free}}(T)=\frac{\lambda_{T}^{3}}{2V}\left\{
Tr_{2}[\text{exp}(-\beta H)]-Tr_{2}[\text{exp}(-\beta H_{\text{free}%
})]\right\}  .
\end{equation}
As before we take $a=(50$ MeV$)^{-1}$ and $L=8$. \ In Fig. \ref{b2_hop1} we
show results for the standard Hamiltonian lattice action at $T=1$.$0$ MeV.
\ \ For comparison we show results for the standard Euclidean lattice action,
the continuum limit, and the standard Hamiltonian lattice action with Galilean
invariance imposed by hand. \ \ We impose Galilean invariance by computing the
spectrum of $H$ in the rest frame. \ We then boost the result for nonzero
total momentum $\vec{p}_{s}$ using%
\begin{equation}
E(\vec{p}_{s})=E(\vec{0})+\frac{\vec{p}_{s}^{2}}{4m}.
\end{equation}%
We see that both the standard Hamiltonian lattice results and standard
Euclidean lattice results deviate from the continuum limit by about the same
amount. \ We also see that the standard Hamiltonian lattice action with
Galilean invariance is almost identical with the continuum limit. \ This
suggests that broken Galilean invariance is in fact responsible for most of
the cutoff error at $T=1.0$ MeV.

\begin{figure}
[ptb]
\begin{center}
\includegraphics[
height=4.5697in,
width=3.2085in,
angle=-90
]%
{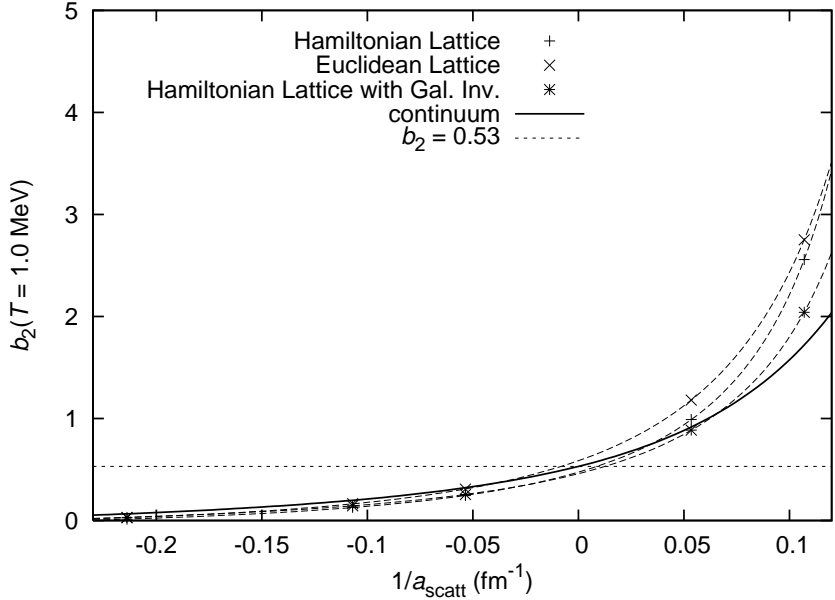}%
\caption{Plot of $b_{2}$ at $T$ = 1.0 MeV for the well-tempered Hamiltonian
lattice action at $T=1$.$0$ MeV. \ For comparison we show results for the
well-tempered Euclidean lattice action, the well-tempered Hamiltonian lattice
action with Galilean invariance imposed by hand, and the continuum limit.}%
\label{b2wt}%
\end{center}
\end{figure}

We show in Fig. \ref{b2wt} results for the $O\left(  a^{2}\right)
$-well-tempered action at $T=1.0$ MeV. \ The four curves shown are for the
$O\left(  a^{2}\right)  $-well-tempered Hamiltonian action, $O\left(
a^{2}\right)  $-well-tempered Euclidean lattice action, the continuum limit,
and the $O\left(  a^{2}\right)  $-well-tempered Hamiltonian lattice action
with Galilean invariance imposed by hand.%
In this case all four curves agree rather well. \ The well-tempered action
clearly preserves Galilean invariance much better than the standard action and
removes most of the cutoff error in both the Hamiltonian and Euclidean lattice formalisms.

\section{Summary and discussion}

In this study we investigated the temperature dependence of lattice
discretization errors in nuclear lattice simulations. \ As a warm-up exercise
we started with the one-dimensional attractive Hubbard model. \ We found that
when the interaction was strongly attractive the dispersion relation for the
two-particle ground state showed significant cutoff errors. \ This cutoff
error could be attributed to the breaking of Galilean invariance. \ The same
problem of strongly-broken Galilean invariance was found in three dimensions
for interacting neutrons with an attractive zero-range potential.

We showed that part of the error due to broken Galilean invariance could be
eliminated by using an $O(a^{2})$-improved lattice kinetic energy. \ The
$O(a^{2})$-improved action includes next-to-nearest neighbor hopping terms in
order to match the single particle dispersion relation $\vec{p}_{s}^{2}/(2m)$
up to terms $O\left(  \left\vert \vec{p}_{s}\right\vert ^{6}\right)  $.
\ While the improved action was better than the standard action, we found even
better results when using an $O(a^{2})$-well-tempered kinetic energy lattice
action. \ The $O(a^{2})$-well-tempered action includes the same
next-to-nearest neighbor hopping terms as the $O(a^{2})$-improved action.
\ However in this case the coefficients of the various terms are adjusted to
match the integral of $\vec{p}_{s}^{2}/(2m)$ over all momenta below the
cutoff$,$%
\begin{equation}%
{\displaystyle\int\limits_{-\Lambda}^{\Lambda}}
{\displaystyle\int\limits_{-\Lambda}^{\Lambda}}
{\displaystyle\int\limits_{-\Lambda}^{\Lambda}}
dp_{x}dp_{y}dp_{z}\;\omega^{\text{(wt1)}}(\vec{p}_{s})=%
{\displaystyle\int\limits_{-\Lambda}^{\Lambda}}
{\displaystyle\int\limits_{-\Lambda}^{\Lambda}}
{\displaystyle\int\limits_{-\Lambda}^{\Lambda}}
dp_{x}dp_{y}dp_{z}\;\frac{\vec{p}_{s}^{2}}{2m}.
\end{equation}

We then performed two separate calculations of the second virial coefficient
$b_{2}(T)$ using the various different lattice actions. \ In the first
calculation we extracted $b_{2}(T)$ in the grand canonical Euclidean lattice
formalism using the virial expansion of the density. \ In the second
calculation we determined $b_{2}(T)$ by a Hamiltonian lattice calculation of
the two-particle partition function. \ In both cases we found that the
$O(a^{2})$-well-tempered lattice action was superior to both the standard
action and $O(a^{2})$-improved lattice action. \ In fact the well-tempered
action reduced the temperature-dependent cutoff errors as much as the
non-local action favored in \cite{Bulgac:2005a}. \ This non-local action
corresponds with the quadratic dispersion relation $\omega^{(\text{quad})}$.
\ However the $O(a^{2})$-well-tempered lattice action has the advantage of
being a local action. \ Therefore it can be implemented in most Monte Carlo
lattice algorithms without increasing the computational scaling.

The well-tempered action is a simple way to reduce lattice errors at nonzero
temperature. \ While the discussion here has focused on zero-range pionless
effective field theory, it seems clear that the well-tempered action fixes the
problem of strongly-broken Galilean invariance quite generally. \ With this
increase in accuracy it should now be possible to perform an accurate lattice
calculation of the third virial coefficient $b_{3}(T)$. \ $b_{3}(T)$ was
recently calculated for two-component fermions in limit of zero effective
range and infinite scattering length \cite{Rupak:2006pu}, and this calculation
can now be checked on the lattice.

We can organize the various kinetic energy actions introduced here somewhat
more systematically. \ In the Hamiltonian lattice formalism let%
\begin{equation}
K_{j-\text{hop}}=\frac{1}{2m}\sum_{\vec{n}_{s},\hat{l}_{s},i=\uparrow
,\downarrow}\left[  a_{i}^{\dagger}(\vec{n}_{s})a_{i}(\vec{n}_{s}+j\hat{l}%
_{s})+a_{i}^{\dagger}(\vec{n}_{s})a_{i}(\vec{n}_{s}-j\hat{l}_{s})\right]
\end{equation}
for integers $j\geq0$. \ In the same way in the Euclidean lattice formalism
let%
\begin{equation}
S_{j-\text{hop}}=\frac{1}{2m}\sum_{\vec{n},\hat{l}_{s},i=\uparrow,\downarrow
}\left[  c_{i}^{\ast}(\vec{n})c_{i}(\vec{n}+j\hat{l}_{s})+c_{i}^{\ast}(\vec
{n})c_{i}(\vec{n}-j\hat{l}_{s})\right]
\end{equation}
for integers $j\geq0$. \ For any chosen lattice action $K^{(n)}$ or
$S_{\text{kinetic}}^{(n)}$ we assign a set of hopping coefficients
$v_{j}^{(n)}$ such that%
\begin{equation}
K^{(n)}=\sum_{j=0,1,2,\cdots}(-1)^{j}v_{j}^{(n)}K_{j\text{-hop}}%
\end{equation}
or%
\begin{equation}
S_{\text{kinetic}}^{(n)}=\sum_{j=0,1,2,\cdots}(-1)^{j}v_{j}^{(n)}%
S_{j\text{-hop}}.
\end{equation}
Then the corresponding single-particle dispersion relation is%
\begin{equation}
\omega^{(n)}(2\pi\vec{k}_{s}/L)=\frac{1}{m}\sum_{j=0,1,2,\cdots}\sum
_{l_{s}=x,y,z}(-1)^{j}v_{j}^{(n)}\cos\left(  \frac{2\pi jk_{l_{s}}}{L}\right)
.
\end{equation}

The $O(a^{4})$-improved action is defined so that its dispersion relation
$\omega^{(2)}(\vec{p}_{s})$ agrees with $\vec{p}_{s}^{2}/(2m)$ up to terms
$O\left(  \left\vert \vec{p}_{s}\right\vert ^{8}\right)  $. \ The $O(a^{4}%
)$-well-tempered action is defined so that its dispersion relation
$\omega^{(\text{wt}2)}(\vec{p}_{s})$ satisfies%
\begin{equation}
\omega^{(\text{wt2})}(2\pi\vec{k}/L)=\omega^{(\text{0})}(2\pi\vec{k}%
/L)+\omega^{(\text{1})}(2\pi\vec{k}/L)+s\left[  \omega^{(\text{2})}(2\pi
\vec{k}/L)-\omega^{(\text{1})}(2\pi\vec{k}/L)\right]  ,
\end{equation}%
\begin{equation}%
{\displaystyle\int\limits_{-\Lambda}^{\Lambda}}
{\displaystyle\int\limits_{-\Lambda}^{\Lambda}}
{\displaystyle\int\limits_{-\Lambda}^{\Lambda}}
dp_{x}dp_{y}dp_{z}\;\omega^{\text{(wt2)}}(\vec{p}_{s})=%
{\displaystyle\int\limits_{-\Lambda}^{\Lambda}}
{\displaystyle\int\limits_{-\Lambda}^{\Lambda}}
{\displaystyle\int\limits_{-\Lambda}^{\Lambda}}
dp_{x}dp_{y}dp_{z}\;\frac{\vec{p}_{s}^{2}}{2m}.
\end{equation}
The generalization to higher-order actions is straightforward. \ The hopping
coefficients for the various actions up to $O(a^{4})$ are shown in Table 1.%
\[%
\genfrac{}{}{0pt}{}{%
\begin{tabular}
[c]{|c|c|c|c|c|c|}\hline
& standard & $O(a^{2})$-improved & $O(a^{2})$-well-tempered & $O(a^{4}%
)$-improved & $O(a^{4})$-well-tempered\\\hline
$v_{0}$ & $1$ & $\frac{5}{4}$ & $\frac{\pi^{2}}{6}$ & $\frac{49}{36}$ &
$\frac{\pi^{2}}{6}$\\\hline
$v_{1}$ & $1$ & $\frac{4}{3}$ & $\frac{2\pi^{2}}{9}-\frac{1}{3}$ & $\frac
{3}{2}$ & $\frac{\pi^{2}}{4}-\frac{13}{24}$\\\hline
$v_{2}$ & $0$ & $\frac{1}{12}$ & $\frac{\pi^{2}}{18}-\frac{1}{3}$ & $\frac
{3}{20}$ & $\frac{\pi^{2}}{10}-\frac{2}{3}$\\\hline
$v_{3}$ & $0$ & $0$ & $0$ & $\frac{1}{90}$ & $\frac{\pi^{2}}{60}-\frac{1}{8}%
$\\\hline
\end{tabular}
}{\text{Table 1: \ Hopping coefficients for kinetic energy lattice actions up
to }O(a^{4})}%
\]
We note that $v_{0}$ for the well-tempered action is $\pi^{2}/6$ for all
orders$.$ \ This is because the integral of $\cos\left(  2\pi jk_{l_{s}%
}/L\right)  $ vanishes for $j\neq0$ and so only the $v_{0}$ term survives in
the integral over momenta. \ These higher-order well-tempered actions may be
useful if we wish to use the same lattice action for high-accuracy
nucleon-nucleon scattering phase shifts and many-body simulations at nonzero temperature.

\section{Acknowledgements}

This work is supported in part by DOE grant DE-FG02-03ER41260.

\bibliographystyle{apsrev}
\bibliography{NuclearMatter}

\end{document}